\documentclass[aps,pre,twocolumn,groupedaddress,superscriptaddress,showpacs]{revtex4-1}
\usepackage{epsfig,amssymb,amsmath,graphicx,subfigure,hyperref}
\usepackage{tabularx}
\usepackage{float}
\usepackage{setspace}
\usepackage{color}

\usepackage{array} 
\newcommand{\PreserveBackslash}[1]{\let\temp=\\#1\let\\=\temp} \newcolumntype{C}[1]{>{\PreserveBackslash\centering}p{#1}} \newcolumntype{R}[1]{>{\PreserveBackslash\raggedleft}p{#1}} \newcolumntype{L}[1]{>{\PreserveBackslash\raggedright}p{#1}} 
\usepackage{textcomp}

\begin{document}

\title{Shapes within shapes: how particles arrange inside a cavity}

\author{Duanduan Wan}
\affiliation{Department of Chemical Engineering, University of Michigan, Ann Arbor, Michigan 48109, USA}
\author{Sharon C. Glotzer}
\email[E-mail: ]{sglotzer@umich.edu}
\affiliation{Department of Chemical Engineering, University of Michigan, Ann Arbor, Michigan 48109, USA}
\affiliation{Department of Materials Science and Engineering and Biointerfaces Institute, University of Michigan, Ann Arbor, Michigan 48109, USA}

\date{\today}
\begin{abstract}
We calculate the configurational entropy of hard particles confined in a cavity using Monte Carlo integration. Multiple combinations of particle and cavity shapes are considered. For small numbers of particles $N$, we show that the entropy decreases monotonically with increasing cavity aspect ratio, regardless of particle shape. As $N$ increases, we find ordered regions of high and low particle density, with the highest density near the boundary for all particle and cavity shape combinations. Our findings provide insights relevant to engineering particles in confined spaces, entropic barriers, and systems with depletion interactions.
\end{abstract}

\maketitle

\section{Introduction}
DNA packaging in viral capsids \cite{Marenduzzo2010}, macromolecular crowding in the cell \cite{Ellis2001}, blood clotting \cite{Cines2014} and pattern formation in biological structures \cite{Hayashi2004} are all examples of important problems involving objects confined inside a cavity. Systems of hard particles confined within cavities  \cite{Manoharan2015,Kamien2007} are of particular relevance to growing colloidal crystals under confinement \cite{Velev2000, Teich2016, Boles2016}, liquid crystals in droplets \cite{Roij2000,Heras2005,Galanis2010,Chen2013}, and entropic barriers for particle transport \cite{Burada2008, Reguera2012}. They also serve as models of colloids with depletants \cite{Asakura1954, Asakura1958, Crocker1999, Zhao2007, Zhao2008} (e.g., Fig.~\ref{shape}(a)), and are relevant, e.g., to the physical situation where particles are confined within walls with shape undulations (Fig.~\ref{shape}(b)).

Although entropy is known to play an important role in these systems, how entropy depends on the shapes of both the particles and the confining cavities, and how that dependence is manifested in the spatial distribution of particles within cavities, are relatively unexplored. Here we use Monte Carlo (MC) integration to calculate the entropy of 2D particles in 2D cavities. We consider the question: if $N$ hard particles are confined within a cavity of fixed area, what combination of particle and cavity shapes maximizes the entropy of the system? We consider only configurational entropy (entropy associated with particle positions and orientations), which for hard particles is given by the accessible free volume within the cavity; this quantity is easily calculated via MC integration. Although we consider only 2D systems here, our results can be generalized to higher dimension.

\section{Methods}
\begin{figure}[h]
\centering 
\includegraphics[width=3.3in]{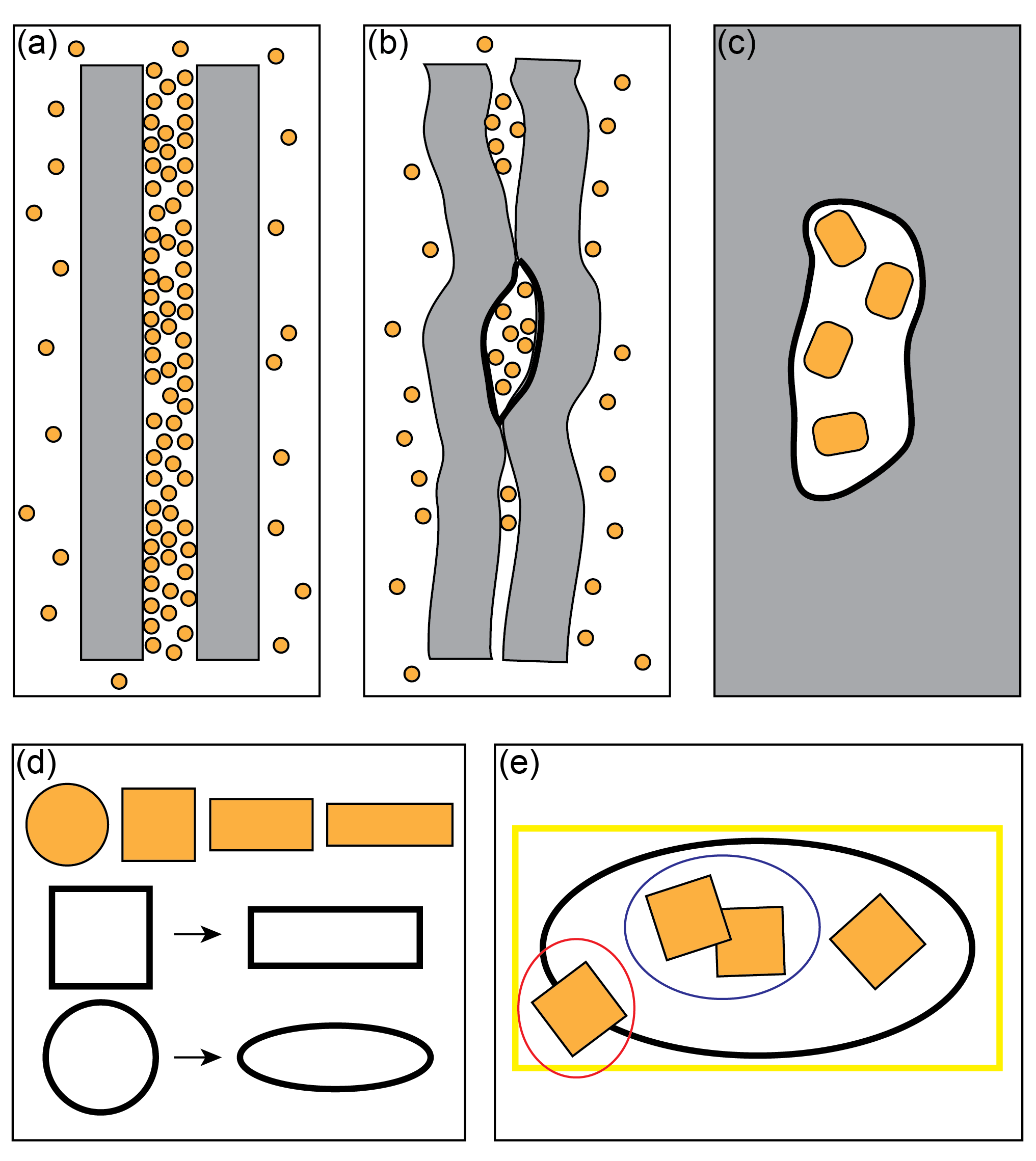}
\caption{(Color online) (a) Cartoon of depletants (small particles, orange) forming layered structures between two aligned plates (gray). (b) Cavity (thick black line indicates the boundary) formed due to shape undulation. (c) An example of small particles confined in a cavity (both particles and cavity can assume arbitrary shapes). (d) Four particle shapes (orange) considered in this work: disk, square, and rectangle with aspect ratio 2 and 3; two cavity shape groups (black lines): rectangular and elliptical, both changing from aspect ratio 1 to 3. (e) An elliptical cavity (black) with aspect ratio 2 and $A_{cav}=16$ in a box (yellow). Particles (orange) are of unit area. Two failure situations: Particles partly outside the cavity boundary (red circle) and overlapping with a previous particle (blue circle). }
\label{shape}
\end{figure}

\begin{figure}[h]
\centering 
\includegraphics[width=3.3in]{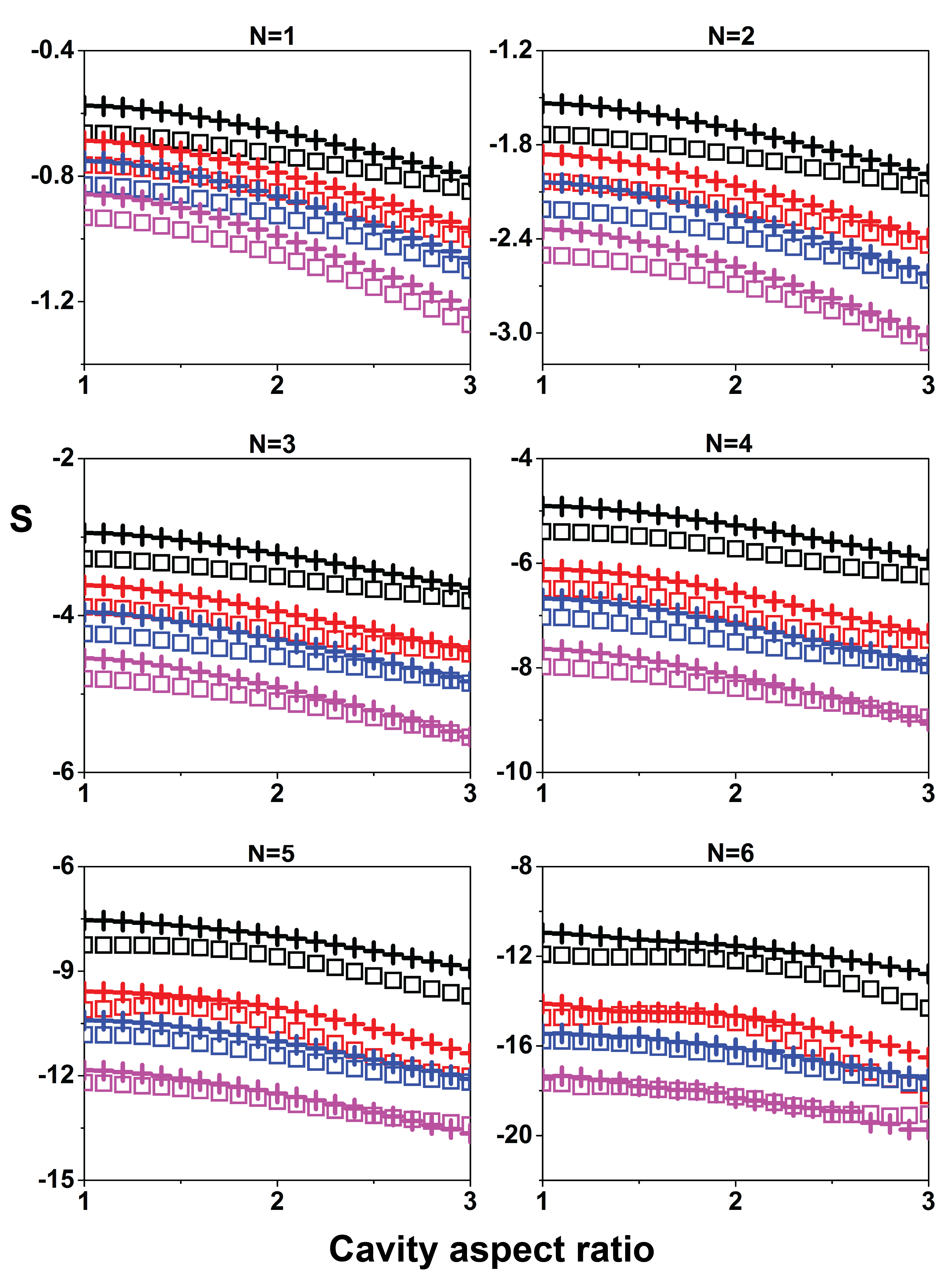} 
\caption{(Color online) Entropy as a function of cavity aspect ratio. Square symbols represent rectangular cavities and plus signs represent elliptical cavities. Black, red, blue and magenta represent disks, squares, and rectangles with aspect ratio 2 and 3, respectively.  Errors are smaller than the size of the marker.} 
\label{entropy}
\end{figure}

Particles and cavities can assume an infinity of shapes depending on the context (see, e.g., Fig.~\ref{shape}(c)). We restrict our study to four simple particle shapes: disks, squares, and rectangles with aspect ratio 2 and 3, as shown in Fig.~\ref{shape}(d). We consider two groups of cavity shapes: rectangular and elliptical, and allow their aspect ratio to change continuously from 1 to 3 (Fig.~\ref{shape}(d)). The free volume available to particles inside the cavities are calculated using MC integration. To perform the integration for a selected particle/cavity shape combination, we first placed the cavity inside a box, as shown in Fig.~\ref{shape}(e). We then placed a particle into the box with its position and orientation randomly selected. If the particle lies completely inside the cavity, we randomly placed a second particle. If the second particle is also inside the cavity boundary and does not overlap with the first one, we placed a third particle. This procedure continued until one of two failure situations shown in Fig.~\ref{shape}(e) occurred. We used the overlap check algorithm from the HOOMD-Blue HPMC module \cite{hoomdblue, Glaser2015} for particle-particle overlap checks for rectangles\cite{Anderson2017}. Overlap checks for disks inside an elliptical cavity was performed using the method in Ref.~\cite{Etayo2006}. Other overlap checks, e.g., two disks, a disk inside a rectangular cavity, or a rectangular particle inside a rectangular or an elliptical cavity, are straightforward. We fixed the particle area at $A_{par}=1$ and cavity area at $A_{cav}=16$. We ran $n_{tot}=5\times 10^{10}$ trials for each particle/cavity shape combination, and counted the number of successful trials $n_{s}$ for each particle number $N$. 

The configurational free volume $V(N)$ of $N$ particles is:
\begin{equation}
V(N)=\frac{n_{s}(N)}{n_{tot}}\times A_{box}^{N}\times (2\pi)^{N}
\label{V}.
\end{equation}
Here $A_{box}$ is the area of the box, the value of which does not affect the free-volume calculation as long as the cavity is inside the box. In practice, we chose the box slightly larger than the cavity to achieve a high acceptance probability. The $(2\pi)^{N}$ factor accounts for particle rotation. The entropy of the particles relative to that of an ideal gas is:
\begin{equation}
S(N)= k_{B} \mbox{ln}\frac{V(N)}{V_{0}(N)},
\label{S}
\end{equation}
where $k_{B}$ is Boltzmann's constant (we set $k_{B}=1$), $V_{0}(N)$ is for normalization and $V_{0}(N)= A_{cav}^{N}\times (2\pi)^{N}$ is the free volume of $N$ ideal gas particles in a cavity (available volume in the phase space) . This normalization makes $S(N)$ independent of the length scale of the system. As $S(N)=0$ for an ideal gas, the entropies we calculate here are negative.

\section{Results and discussion}

\begin{figure}
\centering 
\includegraphics[width=3.3in]{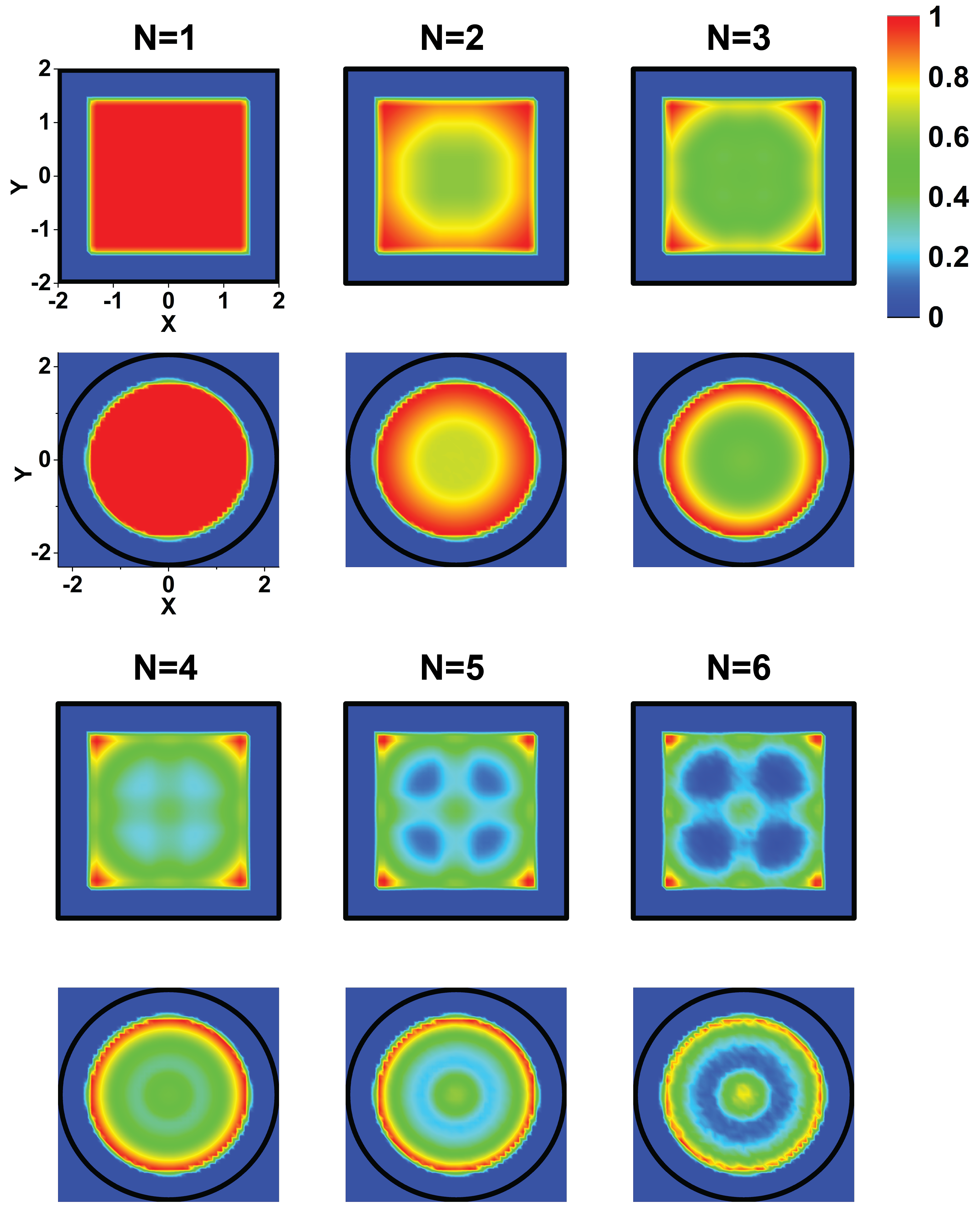} 
\caption{(Color online) Probability distribution of successful trials for disks inside a square cavity (upper) and a circular cavity (lower). Black lines indicate the cavity boundary. Each figure is normalized by the highest probability for that particular case.} 
\label{probability}
\end{figure}

\begin{figure}
\includegraphics[width=3.3in]{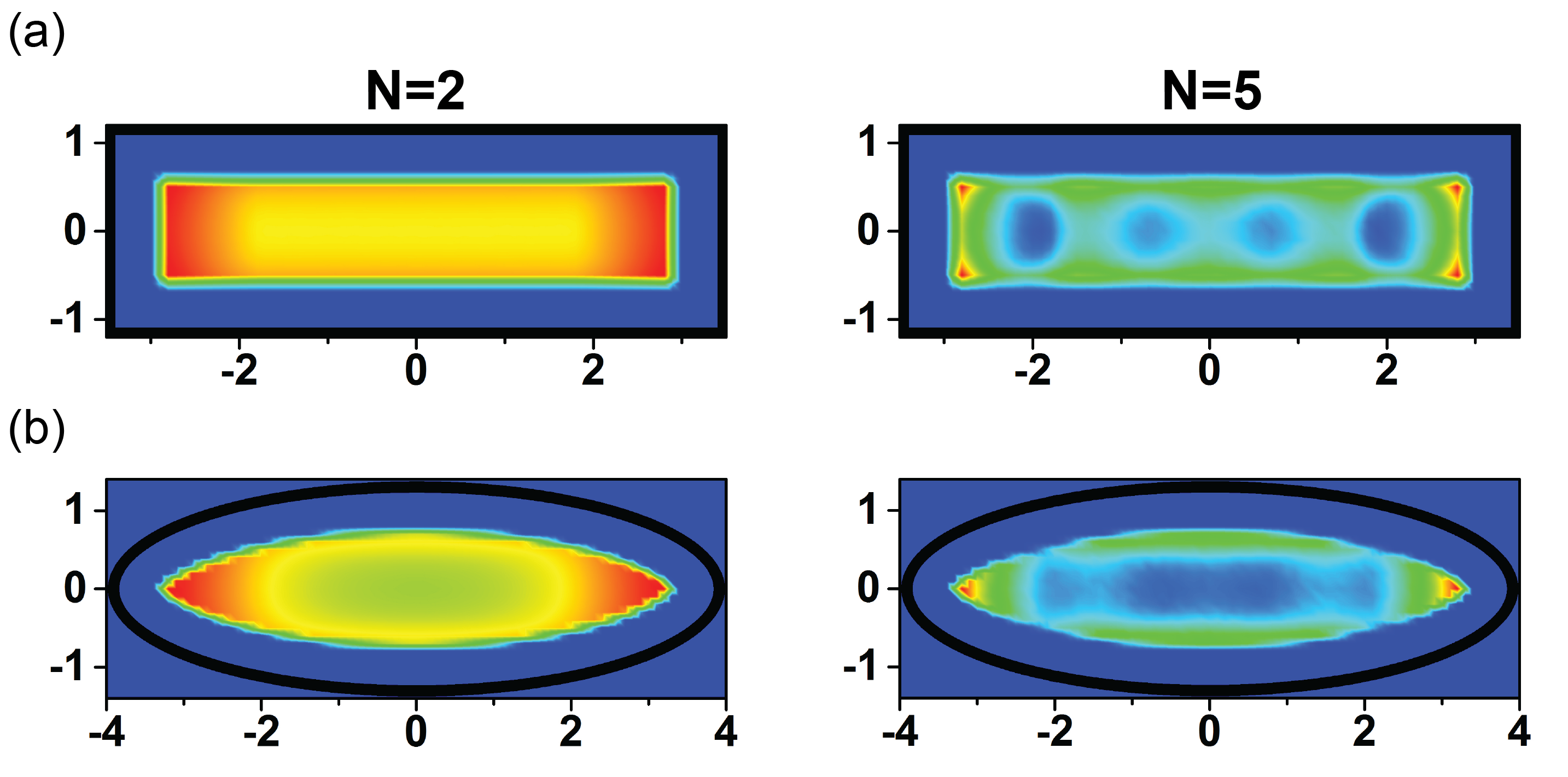} 
\caption{(Color online) Probability distribution for disks inside a rectangular cavity (a) and an elliptical cavity (b), with aspect ratio 3. Black lines indicate the cavity boundary. Color bar is the same as that in Fig.~\ref{probability}.} 
\label{aspect_ratio}
\end{figure}

\begin{figure}
\centering 
\includegraphics[width=2.8in]{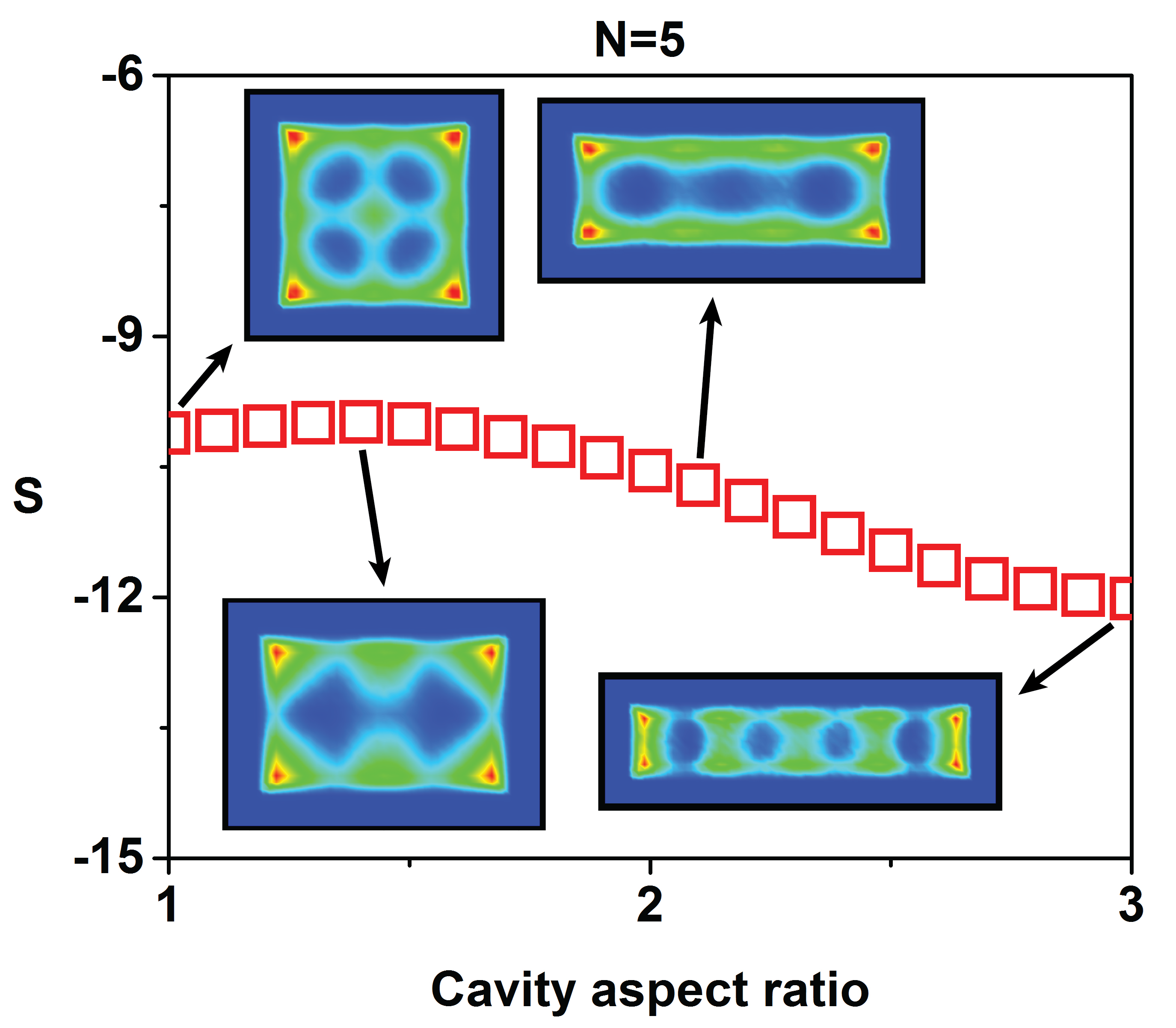} 
\caption{(Color online) Entropy of $N=5$ square particles inside a rectangular cavity, same as that in Fig.~\ref{entropy}. Insets: Probability distributions. Black lines indicate the cavity boundary. Color bar is the same as that in Fig.~\ref{probability}. } 
\label{monotonicity}
\end{figure}

Entropy as a function of cavity aspect ratio is shown in Fig.~\ref{entropy} for particle numbers $N=1$ to $6$. Trials that achieved $N=7, 8$ were rare and thus the errors are relatively large. No trials achieved $N \ge 9$. We summarize several features of Fig.~\ref{entropy}. First, in all cases the entropy decreases monotonically with increasing cavity aspect ratio for $N \le 4$ particles. In the case of square particles inside a rectangular cavity, non-monotonicity appears for $N=5, 6$. Second, the entropy is larger for elliptical cavities compared to their rectangular counterparts for small to intermediate cavity aspect ratio. Sufficiently large aspect ratios are not interesting because ultimately all the curves will decrease to zero. Third, for $N\le 5$ particles in cavities of matching shape, we find $S_{disk} > S_{square} > S_{rect 2} > S_{rect 3}$. For $N=6$, the entropy is smaller for square particles than for aspect ratio 2 rectangles when placed in rectangular cavities of aspect ratios ranging from about $2.7$ to $3$.  
 
We next explore the particle probability distribution (local density). To show the probability distribution, we divide the $xy$ plane into $0.1\times0.1$ grids. For a given $N$, the probability $P(x,y)$ at a grid point $(x,y)$ is
\begin{equation}
P(x,y)=\frac{n_{g}(x,y)}{n_{s}},
\label{plot}
\end{equation}
where $n_{g}(x,y)$ is, among the $n_{s}$ successful trials, the number of times the center of mass of a particle appears at grid point $(x,y)$, i.e., inside the region $(x-0.05, x+0.05), (y-0.05, y+0.05)$. We plot the two largest entropy combinations - disks in circular and square cavities - in Fig.~\ref{probability}. From Fig.~\ref{probability}, disks inside a square cavity, we see that for $N=1$, the density is uniform except at the boundary where it decreases abruptly to zero. In contrast, for other particle shapes, the density is nonuniform due to the breaking of rotational symmetry (see Appendix \ref{app_A} for an example). For $N > 1$ the density begins to increase in the four corners of the square cavity compared to elsewhere in the cavity. This increasing tendency for disks to be found in the corners of a square boundary arises because such configurations increases the amount of accessible space for other disks.  The density enhancement in the corners sharpens as $N$ increases, and we observe ordered regions of high (green and brighter areas in $N=6$ figure in Fig.~\ref{probability}) and low (blue areas) density. 

A similar picture emerges in the case of a circular cavity (Fig.~\ref{probability}, disks inside a circular cavity), where the local density is isotropic and alternates bright and dark in the radial direction. Similarly, when the cavity aspect ratio changes (Fig.~\ref{aspect_ratio}), the highest density appears near the boundary and crystalline order emerges at large $N$. The specific crystalline pattern depends on both the cavity shape and the particle shape. Fig.~\ref{monotonicity} plots the case of $N=5$ square particles inside a rectangular cavity (same as that in Fig.~\ref{entropy}), where we observe non-monotonicity in the entropy. Insets of Fig.~\ref{monotonicity} show the spatially dependent density at selected cavity aspect ratios. The patterns look similar to those observed for disks, but differ in detail. For example, for large $N$, two squares tend to align their edges to increase the accessible free volume \cite{Anders2014PNAS, Anders2014, Manoharan2015}. The configurational entropies and local densities calculated here consider all possible configurations, including those that may be kinetically unrealizable in practice. See Ref.~\cite{Carlsson2012} for a study on the topology of configuration spaces of hard disks confined within a square, and see also e.g., Refs.~\cite{Torquato2010, Donev2004} about jammed states.

Regions of high and low density arise because the cavity breaks the translational invariance, e.g., a rectangular cavity breaks translational invariance in both $x$ and $y$ directions, thus crystalline order forms in both directions. A circular cavity, however,  breaks translational invariance only in the radial direction, thus the probability distribution remains uniform in the angular direction.  This may be better seen from a simplified situation of hard disks in a 1D cavity of length $L$, for which we both numerically and analytically calculated the free volume and local density (see Appendix \ref{app_B} ).  The density is periodic, with the highest values at the boundary and decreasing values away from the boundary. This periodicity in amplitude is large when the particle number is high. The number of peaks is the same as the particle number. The uniform distribution is recovered in the $L \rightarrow \infty$ limit or in the case of periodic boundary conditions. Note that the density in a rectangular cavity is not the product of the probability distributions of two 1D cavities with the corresponding length and width.

\section{Conclusion}
We calculated the configurational entropy for various particle/cavity shape combinations and a range of particle numbers. Our finding that the highest particle density appears at the boundary and is oscillatory is relevant to the layering of particles at walls in much larger systems. For example, 2D hard rectangles confined by parallel walls \cite{Triplett2008,Geigenfeind2015}, small depletants around big particles in depletion systems (e.g., Ref.~\cite{Crocker1999,Roth2000} and references therein), and molecules in a liquid film confined between two surfaces \cite{Israelachvili2011}, all exhibit density profiles with similar oscillation and decay. Here we see that the tendency to form layers is present already for small system size. Our findings may give clues to engineering particles in a confined environment, entropic barriers and systems with depletion interactions, and contribute to a better understanding of systems concerned with objects in geometrical confinement, subjected to more complex interaction besides the hard core potential.

\section{Acknowledgements}
We thank Jens Glaser, Meng Xiao, Michael Engel, Randall Kamien, Erin Teich and Paul Dodd for helpful discussions. This work was supported by a Simons Investigator award from the Simons Foundation to Sharon Glotzer. Computational resources and services were supported by Advanced Research Computing at the University of Michigan, Ann Arbor.

\appendix
\section{A square particle inside a square cavity\label{app_A} }
\begin{figure}
\centering 
\includegraphics[width=3.3in]{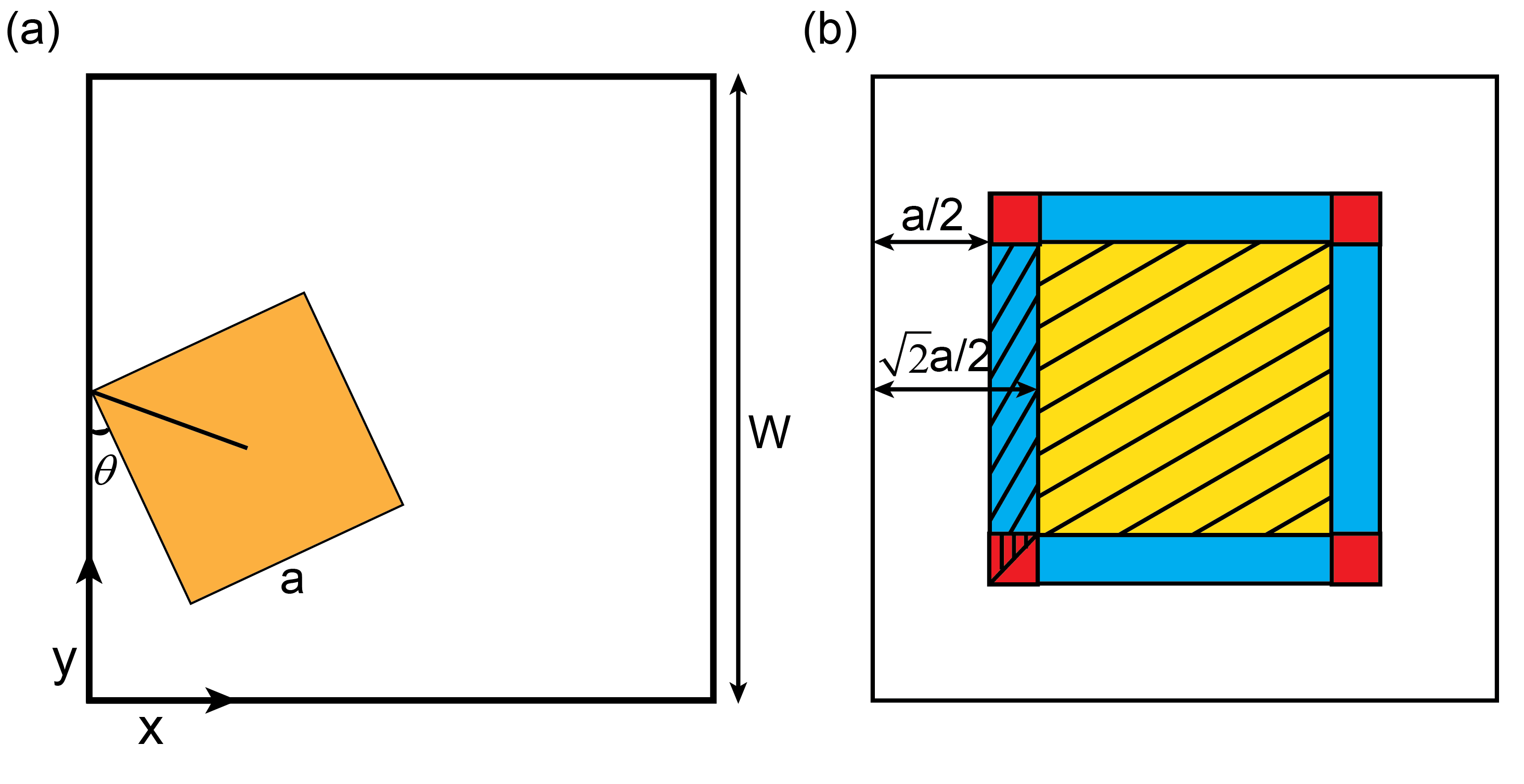} 
\caption{(Color online) (a) A square particle with side length $a$ inside a square cavity with side length $W$. (b) The center of the square particle is inside the colored region: the free-rotating region (yellow) and the constrained rotating regions (blue and red). The shadowed parts are examples of integral regions in the text.} 
\label{square}
\end{figure}

In the following, we take a square particle with side length $a$ inside a square cavity with side length $W$ (Fig.~\ref{square} (a)) as an example and calculate its free volume. In this case, the center of the square particle can exist anywhere in the colored region in Fig.~\ref{square}(b). In the yellow region, the square can rotate freely, and thus the volume is just the area:
\begin{equation}
V_{1}=(W-2 \times \frac{\sqrt{2}}{2}a)^{2}.
\end{equation}
In the blue and red regions, where the particle cannot rotate freely, the free volume at any position is proportional to the angle through which it can rotate. In the shadowed blue region, the angle $\theta$ between one side of the particle and the cavity boundary is given by $\theta = \arcsin(\frac{x}{\sqrt{2}a/2})- \frac{\pi}{4}$ and the free volume is
\begin{equation}
V_{2}=\int_{\frac{a}{2}}^{\frac{\sqrt{2}}{2}a}dx \int_{\frac{\sqrt{2}}{2}a}^{W-\frac{\sqrt{2}}{2}a} dy \, \frac{ [\arcsin(\frac{x}{\sqrt{2}a/2})- \frac{\pi}{4}] \times 2 \times 4}{2 \pi},
\end{equation}
where $x, y$ are the coordinates of the particle center. The $2$ and $4$ in the numerator account for two corners per side and four sides, respectively. Similarly, the volume of the shadowed red region (the lower-left triangle) is
\begin{equation}
V_{3}=\int_{\frac{a}{2}}^{\frac{\sqrt{2}}{2}a}dx \int_{x}^{\frac{\sqrt{2}}{2}a} dy \, \frac{[\arcsin(\frac{x}{\sqrt{2}a/2})- \frac{\pi}{4}] \times 2 \times 4}{2 \pi}.
\end{equation}
The total free volume (besides the $(2\pi)^{N}$ rotational factor) is then
\begin{equation}
V_{tot}=V_{1}+4V_{2}+8V_{3}.
\end{equation}
Taking $W=4$ and $r=\sqrt{1/\pi}$, $V_{tot} \approx 7.4507$. This result has an error of about $0.0004\%$ relative to the simulation result in the main text, which is of the order $1/\sqrt{n_{tot}}$.  


\label{square_cavity}

\section{Disks inside a 1D rectangular cavity \label{app_B} }
\begin{figure}
\centering 
\includegraphics[width=2.8in]{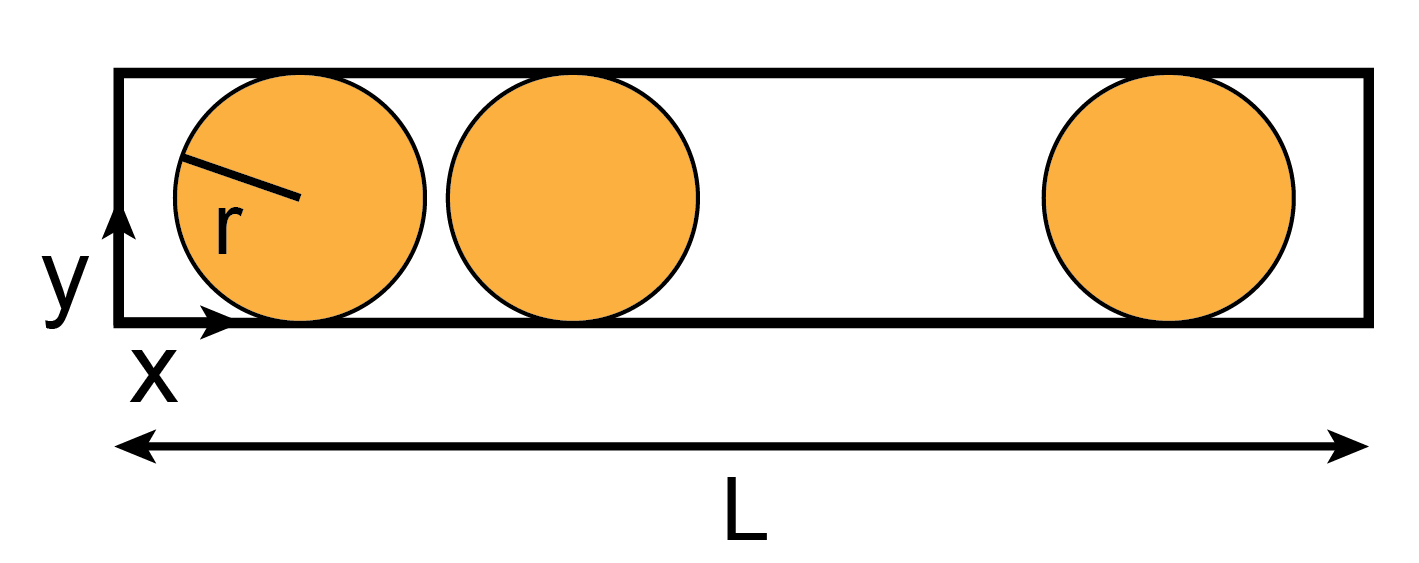} 
\caption{(Color online) Disks inside a 1D cavity. }
\label{1D_cavity}
\end{figure}

\begin{figure*}
\centering 
\includegraphics[width=7in]{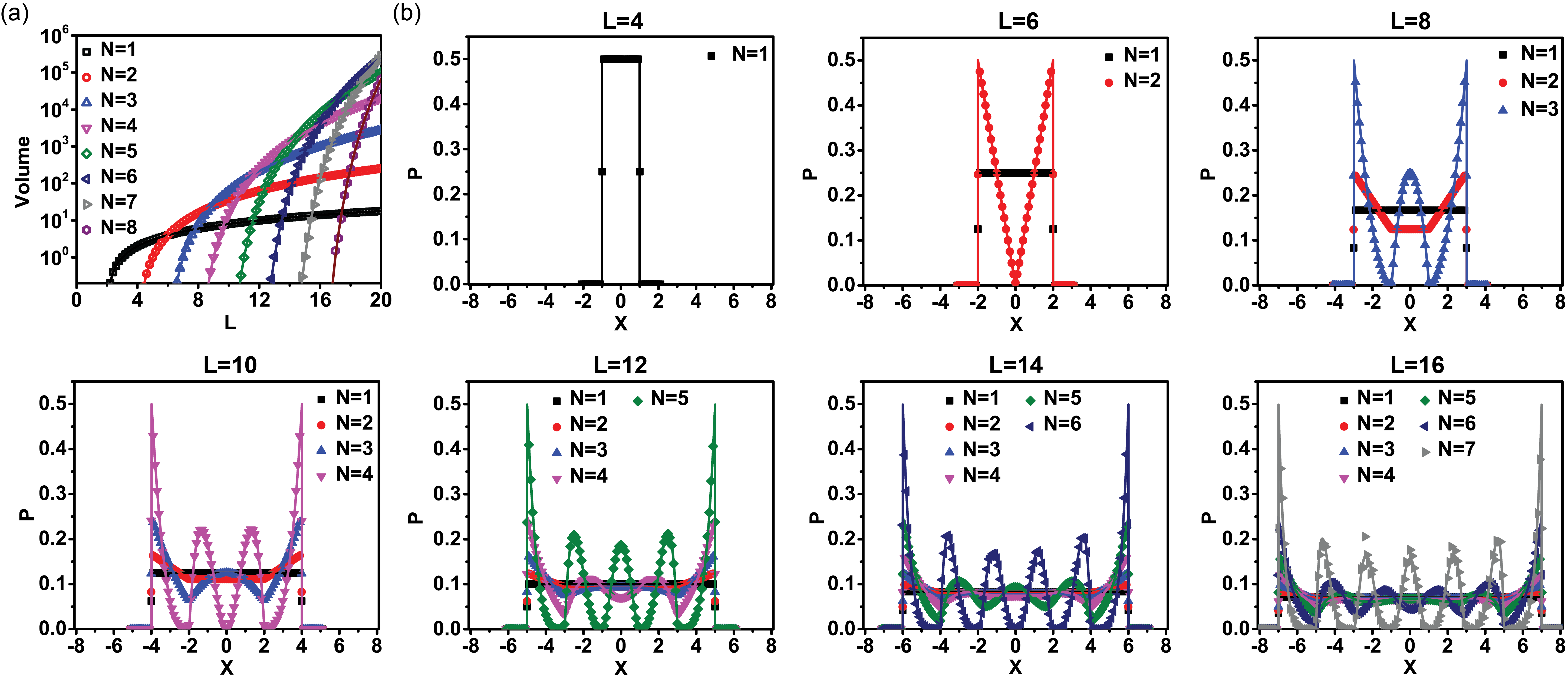} 
\caption{(Color online) (a) Volume of disks with $r=1$ as a function of 1D cavity length from $10^{10}$ tests. Lines are results of Eq.~(\ref{V_1D}). (b) Probability distribution at some $L$ values in (a):4, 6, 8, 10, 12, 14 and 16, respectively. Lines are calculations from Eq.~(\ref{P_1D}), with the cavity centered at $x=0$.} 
\label{disk}
\end{figure*}

The periodic ordering may be better seen from a simplified situation of hard disks in a 1D cavity (Fig.~\ref{1D_cavity}). In this case, the volume (besides the $(2\pi)^{N}$ rotational factor) is
\begin{widetext}
\begin{eqnarray}
V_{1D}(N) = N! \int_{r}^{L-(2N-1)r} dx_{1} \int_{x_{1}+2r}^{L-(2N-3)r} dx_{2} \int_{x_{2}+2r}^{L-(2N-5)r} dx_{3} \cdots   \int_{x_{N-2}+2r}^{L-3r} dx_{N-1} (L-3r- x_{N-1})   = (L-2Nr)^{N}.   \label{V_1D}  \nonumber  \\
\end{eqnarray}
\end{widetext} 
Here $L$ is the cavity length, $r$ is the disk radius, and $x_{i}$ are the coordinates of the centers of the $i$th particle. Eq.~(\ref{V_1D}) is the same as that in the 1D gas of the Tonks model \cite{Tonks1936,Kamien2007}. Fig.~\ref{disk}(a) shows the results from simulation and using Eq.~(\ref{V_1D}). The corresponding probability of finding a disk at position ${x}$ then reads
\begin{widetext}
\begin{eqnarray}
P(x)=\frac{\sum_{m}{{N-1}\choose{m}}(x-r-2mr)^{m}[L-x-r-2(N-1-m)r]^{N-1-m}}{(L-2Nr)^{N}}, 
\label{P_1D}
\end{eqnarray} 
\end{widetext}
for all integers $m$: $m\geq0$ and $N-1/2-(L-x)/2r \leq m \leq (x-r)/2r$, where ${{N-1}\choose{m}}$ is the binomial coefficient and $0 \leq x \leq L$. Fig.~\ref{disk}(b) shows the particle probability distribution (particle density) at selected $L$ values.

\bibliographystyle{apsrev4-1}
\bibliography{cavity_references}

\end{document}